\documentstyle[prl,aps,twocolumn]{revtex}

\begin{document}
\twocolumn[\hsize\textwidth\columnwidth\hsize\csname
  @twocolumnfalse\endcsname
\vspace{-0.5in}
\draft
\title{Lorentz Transformation from Symmetry of Reference Principle}
\author{M. Dima}
\address{Nuclear Science \& Technology Dept.,\\
         Harbin Engineering University, \\
         Harbin,
	 Heilongjiang, CN-150000, China}
%
%
\maketitle
\begin{abstract}
   The Lorentz Transformation is traditionally derived requiring
   the Principle of Relativity and 
   light-speed universality. While the latter can be 
   relaxed, the Principle of Relativity is seen as core to
   the transformation.
   The present letter relaxes
   both statements 
   to the weaker, {\sl Symmetry
   of Reference} Principle.
   Thus the resulting 
 Lorentz transformation and its consequences
 (time dilatation, length contraction)
 are, in turn, effects of how we manage
 space and time.
\end{abstract}
 ] 

 Starting with the 1905 paper of A. Einstein
 in {\sl Ann. Phys.}~\cite{bib:ae}
 the
 Lorentz Transformation has been
 traditionally derived
 based on the Principle of
 Relativity and light-speed
 universality. A number of studies~\cite{bib:lev}
 have shown that light-speed universality
 is not needed - the first such publication (1906)
 being owed to H. Poincar\'{e}~\cite{bib:poin}.
 Group theory expresses the transitivity
 property of relativity (C relative to A, if A to B
 and B to C) in the form
 of the group closure relation, respectively 
 the product of two group elements
 being another element of the group.
 Pure relativity transformations 
 however, cannot form a group on their own,
 needing rotations to ``close" the group.
 As such, the full group is not immediate from
 the Principle of Relativity and
 needs to be specified (in this case a group of 
 transformations invarying the metric: 
 ${\mathbf \Lambda}^\dagger {\mathbf G} {\mathbf \Lambda} = {\mathbf G}$,
 where ${\mathbf \Lambda}$ are the transformations and
 ${\mathbf G}$ the metric). Entering the Lorentz group however,
 is equivalent to admitting light speed invariance.
 In this sense the Principle of Relativity and (indirectly)
 light speed invariance are
 core
 to the Lorentz Transformation~\cite{bib:berz}.
  
  The present letter shows however, that neither statement
 is necessary and that the Lorentz transformation
 stems from the simpler (weaker) Principle of
 Symmetry of reference systems. Further more, the Minkowsky
 metric is not unique in defining relativity.
 There are two possible classes of
 transformations, one invarying the
 Minkowsky and the other the Euclidian metric.
 The ad-hoc terminology of Minkowsky and Euclidian relativities
 will be thus adopted throughout this letter.

  Consider two coordinate systems in motion
  that at some point were at
  rest relative to each other
  and were aligned to have the same
  orientation, offset and
  (Euclidian) space-metric.
  The transformation between such coordinate systems is:
  \begin{equation}
     \bigg(
       \begin{array}{c}
        dt                          \\
        d\vec{x}  
      \end{array}
     \bigg)' \, = \,
     \underbrace{
      \left(
      \begin{array}{cc}
        {\rm scalar}_{_{1 \times 1}}        \, \, \, \, \,  &
	{\rm vector}_{_{1 \times 3}}                          \\
        {\rm vector}_{_{3 \times 1}}       \, \, \, \, \,  &
	{\rm \bf tensor}_{_{3 \times 3}}  
      \end{array}
      \right)
      }_{   {\mathbf \Lambda}}
      \bigg(
       \begin{array}{c}
        dt                          \\
        d\vec{x}  
      \end{array}
     \bigg)
  \end{equation}
 where the dimensions of the objects involved is given
 by the subscripts.  
  In general the 
  transformation should be
  an integral, non-linear transformation, however
 general considerations about space-time limit the range of
 possible transformations to constant linear transformations:
  \begin{enumerate}
  \item{{\sl locality} - implies that the transformation must 
                         be point-to-point;}
  \item{{\sl homogeneity} - implies that the transformation must 
                         not depend on the relocation of the coordinate 
                         system, hence linear;}
  \item{{\sl isotropy} - implies that the mathematical objects in
                        the transformation cannot be ``pseudo"-objects,
			since coordinate system space/time inversion
			must not affect the
			transformation. Also, the vectors must be parallel
			to $\vec{v}$ - the relative velocity between
			the systems in causa, as no new direction
			in space can be introduced (isotropy).
			Likewise, the general form of the tensor
			- less a ($\times \vec{v}$) pseudo-tensorial
			part ruled out by isotropy, is
			$\lambda {\mathbf C}_\parallel +
                 \mu {\mathbf C}_\perp$,
 where ${\mathbf C}_\parallel$  selects 
 vector components parallel to $\vec{v}$,
 and ${\mathbf C}_\perp$ components perpendicular to $\vec{v}$.
 The scalars must be functions of $|\vec{v}|$, as there exists no preferred
 direction in space.
}
  \end{enumerate}
 
  The transformation can be thus written as:
  \begin{equation}
    {\mathbf \Lambda} =
      \gamma_v
      \left(
      \begin{array}{cc}
        1                          \, \, \, \, \,  &
	- \vec{v}/c_v^2s                              \\
        - \psi_v \vec{v}             \, \, \, \, \,  &
	\lambda_v {\mathbf C}_\parallel + \mu_v {\mathbf C}_\perp  
      \end{array}
      \right)
  \end{equation}
 with the scalars fulfilling the roles described above and $s$
 a sign factor $s = \pm 1$. For reasons
 evident later the two    
 shall be termed Minkowsky ($s = +1$) and Euclidian relativity
 ($s = -1$). 

  The apparent-velocity of an object moving with $\vec{u}$
  in the base-system is seen in the moving-system as:
  \begin{equation}
      \vec{u} \ominus \vec{v} =
      \frac{-\psi_v \vec{v} + \lambda_v \vec{u}_\parallel 
      + \mu_v \vec{u}_\perp}
           {1- \vec{v}\vec{u}/c_v^2s}
  \end{equation}
 the scalar $c_v$ having units of speed.
 
  The following are evident:
  \begin{enumerate}
  \item{for $\vec{v} \rightarrow 0$ the transformation
        is unitary:
	\begin{equation}
          \lim_{v \rightarrow 0} {\mathbf \Lambda}_{\vec{v}} = {\mathbf 1}
	\end{equation}
	hence $\gamma_v$, $\lambda_v$, $\mu_v$ = 1,}
  \item{$\vec{v} \ominus \vec{0}$ = -
        $\vec{0} \ominus \vec{v}$ 
	$\, \,$
	thus 
	$\, \,$
	$\psi_v$ = 1.}
  \item{$\vec{v} \ominus \vec{v}$ = $\,  \, \, \,\vec{0}$ 
	$\, \, \, \, \, \, \, \, \, \, \, \, $
	thus 
	$\, \, \,$
	$\lambda_v$ = $\psi_v$ = 1.}
  \item{${\mathbf \Lambda}_{-\vec{v}} =
	 {\mathbf \Lambda}_{ \vec{v}}^{-1} $ 
	 $\, \, \, \, \, \, \, \, \, \, \, \, \, $thus:
	\begin{itemize}
	  \item{$\gamma_v = \pm 1 / \sqrt{1- \vec{v}^2/c^2s}$, 
	  the valid sign (+) being determined from $\lim_{v \rightarrow 0}$,}
	  \item{$\mu_v = \pm 1/\gamma_v$,
	  the valid sign (+) being determined from $\lim_{v \rightarrow 0}$.}
	\end{itemize}
	}
  \item{$|\vec{v}_2 \ominus \vec{v}_1|$ =
	$|\vec{v}_1 \ominus \vec{v}_2|$
	$\,$ implies 
	$c_1$ = $c_2$ = $c$ a constant and $s_1$ = $s_2$ = the same sign.  }
  \end{enumerate}
  From the apparent-velocity law, the combined
  speed is:
  \begin{equation}
    |\vec{u} \ominus \vec{v}|^2 =
    \frac{(\vec{u}-\vec{v})^2 - s(\vec{u}\times\vec{v})^2}{(1-\vec{u}\vec{v}/c^2s)^2} 
  \end{equation}
  confirming $v_{\rm lim} \le c$ for $s$ = +1
  at point ({\sl 4}) above.
	   The two cases  
  are somewhat similar for $v \, {\rm or} \, u > c$,  
  but differ significantly for $v \, {\rm and} \, u < c$.

  The transformation is thus now:
  \begin{equation}
    {\mathbf \Lambda} =
      \left(
      \begin{array}{cc}
        \gamma_v                          \, \, \, \, \,  &
	- \gamma_v \vec{v}/c^2s                              \\
        - \gamma_v \vec{v}             \, \, \, \, \,  &
	\gamma_v {\mathbf C}_\parallel + {\mathbf C}_\perp   
      \end{array}
      \right)
  \end{equation}
  respectively
  the well known Lorentz transformation for $s = +1$.
  The meaning of $c$ is related to {\sl causality}:
  for both Minkowsky
  and Euclidian relativities
  the transformed time interval
  versus proper time
  is $dt' = \gamma_v (1 - \vec{v}\vec{u}/c^2s)d\tau$,
  respectively a
  {\sl causal} transformation for $v < c$.
  For $c \rightarrow \infty$ the Galilean transformation is
  recovered.

  The Principle of Relativity (group 
  ``closure" - modulo a
  rotation) has not 
 been used thus far:
  \begin{equation}
    {\mathbf \Lambda}_1 {\mathbf \Lambda}^{-1}_2 = 
    {\mathbf R} {\mathbf \Lambda}_{12}
  \label{eq:12R}
  \end{equation}
 where ${\mathbf \Lambda}_{1,2}$ are two coordinate transformations,
 ${\mathbf \Lambda}_{12}$ the system-1 to system-2 transformation
 and ${\mathbf R}$ an alignment rotation 
 $(\vec{v}_2 \ominus \vec{v}_1) = - {\mathbf R} 
 (\vec{v}_1 \ominus \vec{v}_2)$ that appears 
 due to boost when referencing is done
 via alignment with a (third party) base-system.
    In a strong sense the Principle of Relativity is
 the group ``closure"~\cite{bib:poin}
 relation (\ref{eq:12R}) - satisfied by both Euclidian and
 Minkowsky relativities, however in a weaker form
 it has been used in relations (2), (3) and (5) as the {\sl Symmetry 
 of Reference} Principle. 
  The only isotropic metrics invariant under the transformations
  are the Euclidian ($s = -1$) and the
  Minkowsky ($s = +1$) metric.
  
   Principially tachyons are allowed in both Minkowsky and Euclidian
  relativities, with the notable difference that
  continuous acceleration across the light-cone is possible 
  only in the latter.
   For tachyons 
  time can be seen as running backward if $\vec{u}\vec{v}s > c^2$.
  Since the beginning and end points of a tachyon track
  are not labeled, the tachyon could
  look like a sub-luminous particle. In certain contexts
  the apparent speed of a tachyon
  can even be $v_{\rm tachyon} \ll c$.
  Another interesting aspect of tachyons is their
  apparent charge $\rho' = \gamma_v \rho (1-\vec{v}\vec{u}/c^2s)$,
  of opposite sign to $\rho$ when the apparent time of the
  tachyon is
  running backward. Apparent time-reversal
  couples well with {\sl charge-conjugation}, which is supposed to
  turn {\sl all} currents into anti-currents, rather than each
  one individually - with the $\gamma_{_{\rm Dirac}}$ matrices
  of its own quantum space. With
  respect to other charge-conjugation representations,
  time-reversal ($\gamma^0 \gamma^5$ in the 
  Dirac theory) is Dirac-representation independent and
  space-coordinate independent.  
  Minkowsky relativity has the problem of imaginary
  $\gamma_v = i/\sqrt{v^2/c^2-1}$ for
  $v > c$, however tachyon dynamics is
  expressed
  in both  
  Minkowsky and Euclidian relativity with $\gamma^2$:
  \begin{equation}
     m_{_0}\vec{a} =  
     \bigg(\frac{1}{\gamma^2}{\mathbf C}_\parallel + {\mathbf C}_\perp\bigg)
     \frac{d\vec{p}}{dt}
  \end{equation}
  the acceleration parallel to the force diverging
  with increasing speed. In 
  Minkowsky relativity however,
  as particles approach the light-cone (from above, or below) acceleration
  in the direction of the force is annulled.

 $\,$ \\
 \indent
 I am thankful for the hospitality during completion of this
 work to the Nuclear Science \& Technology Dept.
 of the Harbin Engineering University - China
 under a Foreign Expert grant.


\begin{references}
 
 \bibitem{bib:ae} A. Einstein, {\sl Ann. Phys.} {\bf 17},
  891 (1905).
 \bibitem{bib:lev} Examples in 
 J.M. Levy-Leblond, {\sl Am. J. Phys.} {\bf 44},
  271 (1976).
 \bibitem{bib:poin} H. Poincar\'{e}, 
 {\sl Rend. Circ. Mat. Palermo},  {\bf 21},
  129 (1906).
  \bibitem{bib:berz} V. Berzi, V. Gorini, 
 {\sl J. Math. Phys.} {\bf 44},
  271 (1976).
  \bibitem{bib:dima} M. Dima,  
 {\sl JETP Lett.} {\bf 72},
  541 (2000).



 \end{references}
\end{document}